\documentclass[10pt,conference]{IEEEtran} 
\usepackage{cite}
\usepackage{graphicx} 
\usepackage{xspace}
\usepackage{tikz}
\usepackage{pgfplots} 
\pgfplotsset{compat=1.18}
\usepackage{subcaption}
\usepackage{pgf-pie}
\usepackage{enumitem}
\usepackage{booktabs}
\usepackage{comment}
\usepackage{xurl}
\usepackage[hidelinks]{hyperref}
\newcommand{\eg}{e.g.,\xspace}
\newcommand{\ie}{i.e.,\xspace}
\newcommand{\etc}{etc.\xspace}

\newcommand{\cut}[1]{}

\newcommand{\dotitem}{\noindent$\bullet$\ }

\newcounter{insight}

\newcommand{\ibox}[3]{%
\refstepcounter{insight}%
\vspace{1pt}
\noindent
\framebox[\columnwidth]{%
  \parbox{\dimexpr\columnwidth-2\fboxsep-2\fboxrule\relax}{%
    \textbf{Insight~\theinsight: #2.} #3%
  }%
}%
\label{#1}%
\vspace{1pt}
}

\title{A Large-Scale Comprehensive Measurement of AI-Generated Code in Real-World Repositories}
\author{
\IEEEauthorblockN{
Tianhao Mao\textsuperscript{1},
Dongfang Zhao\textsuperscript{1},
Haixu Tang\textsuperscript{1}
XiaoFeng Wang\textsuperscript{2},
Hang Zhang\textsuperscript{1},
}
\IEEEauthorblockA{
\textsuperscript{1}Luddy School of Informatics, Computing, and Engineering, Indiana University Bloomington \\
\textsuperscript{2}College of Computing and Data Science, Nanyang Technological University\\
\\
Email: tianmao@iu.edu, zhaodo@iu.edu, hatang@iu.edu, xiaofeng.wang@ntu.edu.sg, hz64@iu.edu}
}

\begin{document}

\maketitle
\begin{abstract}
Large language models (LLMs) are rapidly transforming software engineering by enabling developers to generate code ranging from small snippets to entire projects.
As AI-assisted code becomes increasingly integrated into real-world systems, understanding its characteristics and impact is critical.
Existing study on AI-generated code is usually limited in the lab setting with synthetic benchmarks and small-scale coding tasks and covers limited metrics.
AI-assisted code's manifestation in real-world codebases and its differences between human-written one remain unclear.

To close this gap, we perform a first large-scale measurement study of AI-assisted code,
in comparison with the human-written,
in real-world repositories. 
We study a comprehensive set of metrics including both code-level aspects (\eg structural and graph-level complexity, coding style, security quality, \etc) and commit-level characteristics (\eg commit size, frequency, post-commit stability, \etc).

Our results provide new findings and insights:
some contrast previous observations in the lab setting (\eg we conclude that real-world AI-Human differences on code-level metrics are rather small instead of more pronounced),
some extend prior results with finer-grained observations (\eg the variance of security quality across different programming languages),
yet more are presented for the first time on aspects not covered before (\eg code duplication rate, commit size and stability, \etc).
Based on these comprehensive real-world results,
we also discuss the practical implications of AI-assisted programming.

\end{abstract}
\section{Introduction}
\label{sec:intro}


Recent advances in artificial intelligence, particularly large language models (LLMs), are profoundly reshaping modern software engineering practices.
Systems such as OpenAI Codex and Claude Code, along with IDE-integrated assistants like GitHub Copilot and Cursor, have enabled developers to offload a wide spectrum of programming tasks to AI.
These tools are now routinely used, 
from small code snippet completion to whole project synthesis.
As a result, AI-assisted programming is transitioning from an experimental capability to a mainstream development paradigm, fundamentally altering how software is written and maintained.

This paradigm shift raises critical questions about the characteristics and implications of AI-generated code in real-world software systems.
While early studies~\cite{ESEM11323370,ESE2026,ICSE55347.2025.00040,ISSRE11229706,xu-etal-2026-code} have examined LLM-generated code in controlled environments - typically focusing on small, self-contained programming tasks or benchmark datasets - such settings fail to capture the complexity and dynamics of real-world development. 
In practice, AI-generated code is often intertwined with human-written code, iteratively refined, and deployed within large, evolving codebases.
Understanding the properties of such code therefore requires moving beyond lab-based evaluations toward comprehensive empirical studies in real-world repositories.

A key aspect of this understanding lies in analyzing code-level characteristics of AI-generated code. These include traditional software quality and complexity metrics, such as cyclomatic complexity, control-flow and data-flow complexity, language-specific constructs (\eg object-oriented metrics), \etc.
Additionally, defect-related indicators, such as bugs per KLOC, are essential for assessing reliability and maintainability.
Comparing these metrics between real-world AI-involved and human-written code can provide insights into whether AI assistance leads to simpler, more complex, or potentially more error-prone implementations.

Beyond code-level properties, it is equally important to investigate commit-level characeristics of AI-assisted code in comparison to human-written one.
Modern software engineering is collaborative and version-controlled, making commits a natural unit of analysis.
Questions arise as to whether and how AI-assisted commits differ from human commits regarding different aspects:
Do AI-involved commits modify larger portions of code?
Do they accelerate development by increasing commit frequency?
How stable are such commits - do they require more frequent subsequent fixes, revisions, or reverts?
Addressing these questions is crucial for understanding the broader impact of AI on software engineering processes.

However,
relevant work primarily focuses on limited sets of code-level metrics and evaluates LLM-generated outputs in controlled, small-scale settings~\cite{rahman2025,guo2026,SMC10394237}.
To the best of our knowledge,
no prior study exists for a comprehensive, large-scale empirical analysis that jointly examines both code-level and commit-level characteristics between AI and human code,
in a real-world setting.

This paper aims to bridge this gap.
A central challenge in this endeavor is identifying and collecting AI-generated code from real-world repositories, where such information is not explicitly labeled.
To address this, we develop a multi-stage rule-based comment filter and LLM-assisted classifier to identify AI-involved code from self-admitted code comments.
Our pipeline collects a large-scale dataset (largest by far to our knowledge) of likely AI-generated code in real repositories.
Based on it,
we extensively measure multiple code-level and commit-level metrics.

Our study yields new findings and insights. Some of them contradict to the results of previous measurements in less realistic setting, for example, we find that code-level differences (on 50+ metrics) between AI and human code in real-world repositories are much smaller than previously works~\cite{10.1109/ICSE55347.2025.00005,xu-etal-2026-code,rahman2025,ISSRE11229706}, suggesting the developers' careful and controlled AI usage for real-world tasks.
Some extend the previous results, for example, our fine-grained analysis of security issue distribution (\S\ref{sec:res-codeql}) covers more languages than previous work~\cite{ISSRE11229706,SMC10394237,mohsin2024,wang2024} and reveal interesting \emph{language dependent} AI-human differences regarding code quality.
Some of our results are based on new metrics that have not been previously measured,
such as code duplication (\S\ref{sec:res-code-style}) and many commit-level aspects (\S\ref{sec:commit-measure}),
they all bring fresh insights.


In summary, we make the following contributions:

\dotitem We build and will release a large-scale (by far largest to our knowledge) dataset of likely AI-generated code extracted from real-world repositories, enabling various analytical study.

\dotitem Based on the dataset, we perform comprehensive measurement study of both code-level and commit-level characteristics, covering an extensive set of metrics, comparing likely AI-generated and human-written code.

\dotitem We distill multiple insights from our measurement findings regarding AI-assisted programming in practice and discuss potential implications for stakeholders including programmers and AI vendors.
\section{Related Works}
\label{sec:rel}
Most prior work evaluates LLM-generated code in prompt-based or benchmark-style settings, where researchers design programming tasks, ask LLMs to generate code, and then analyze the resulting outputs. For example, \cite{ESEM11323370} studied differences in maintainability and reliability between LLM-generated and human-written code, while \cite{ESE2026} examined performance and efficiency, and \cite{ICSE55347.2025.00040} focused on call calibration and correctness of LLM-generated code. \cite{xu-etal-2026-code} compared code style differences between human-written and LLM-generated code and further tracked style trends in GitHub open-source projects. Other studies investigated code-level features such as line counts, cyclomatic complexity, nesting depth, and class counts \cite{rahman2025}, while \cite{guo2026} examined lexical properties such as lexical density, \cite{ISSRE11229706} compared AI-generated code and human-written code in security and several code structure metrics mentioned above. However these features were often used mainly for classification or detection purposes, rather than for a systematic characterization of how LLM-generated code differs from human-written code across multiple structural dimensions.

Security is another major focus in prompt-based evaluations of LLM-generated code. Early work conducted exploratory assessments of security to ChatGPT-generated code\cite{SMC10394237}. Subsequent studies expanded the evaluation to multiple models and further examined how model parameters and prompt strategies affect the security of generated code\cite{TETCI10658990}. To enable more systematic assessment, \cite{Siddiq_2024} proposed a unified framework integrates code generation and security analysis, while\cite{mohsin2024} provided more detailed study for impact of prompt engineering on code security. \cite{sajadi2025} studied LLM-based security patching, which also sheds light on the security implications of LLM-generated code. In addition, \cite{wang2024} and \cite{ICSE202310548523} developed datasets and benchmarks for more systematic evaluation of the quality and security of LLM-generated code. Some work has further explored mitigation strategies, such as prompt engineering for safer code generation\cite{IFIP10.1007/978-3-031-92886-4_8} and model fine-tuning to improve the security of generated code\cite{CCS10.1145/3576915.3623175}. Relatedly, several studies have investigated the use of LLMs themselves for vulnerability and bug discovery in code\cite{ndssLinM25,ASE202511334273}. Overall, these studies provide important foundations for understanding the security of LLM-generated code, but they still predominantly evaluate model outputs under controlled settings.

Beyond controlled prompting, a growing body of work has begun to examine LLM-involved code in real-world open-source development. \cite{pearce2022asleep} and \cite{TOSEM10.1145/3716848} studied GitHub Copilot-generated code in real GitHub repositories, focusing on security issues and the ability of LLMs to repair vulnerabilities. \cite{xiao2026} analyzed self-admitted LLM usage in GitHub repositories, examining how generative AI is used in practice and surveying developers about usage policies. Using the AIDev dataset \cite{li2025aidev} of LLM-agent pull requests, \cite{siddiq2026} investigated security-related issues, review interactions, and ecosystem characteristics, while \cite{hasan2026} examined cyclomatic complexity and code quality in the same setting. Other studies focused on specific aspects of AI-involved code in repositories, such as code smells in GitHub Copilot-generated code \cite{ASE2024} and adoption patterns reflected in modified lines and touched files at both commit and file levels for AI-agent-generated code \cite{robbes2026}. These studies move beyond synthetic prompting scenarios and provide valuable evidence about how LLMs are actually used in practice.

Another related line of work studies the detection and measurement of LLM-generated code in the wild. \cite{science.adz9311} measured the usage of LLMs for coding by training a classifier for LLM-generated code, while \cite{ICSE55347.2025.00064,orel-etal-2025-codet,10.1109/ICSE55347.2025.00005}, and \cite{AISec10.1145} also explored methods for detecting LLM-generated code. These studies are closely related because they help identify AI-generated code in large-scale repositories, but their main goal is detection or usage estimation rather than a systematic comparison of the characteristics of real-world LLM-generated code and human-written code.
\section{Measurement Pipeline}
\label{sec:pipeline}

\begin{figure}[!t]
    \centering
    \includegraphics[width=\linewidth]{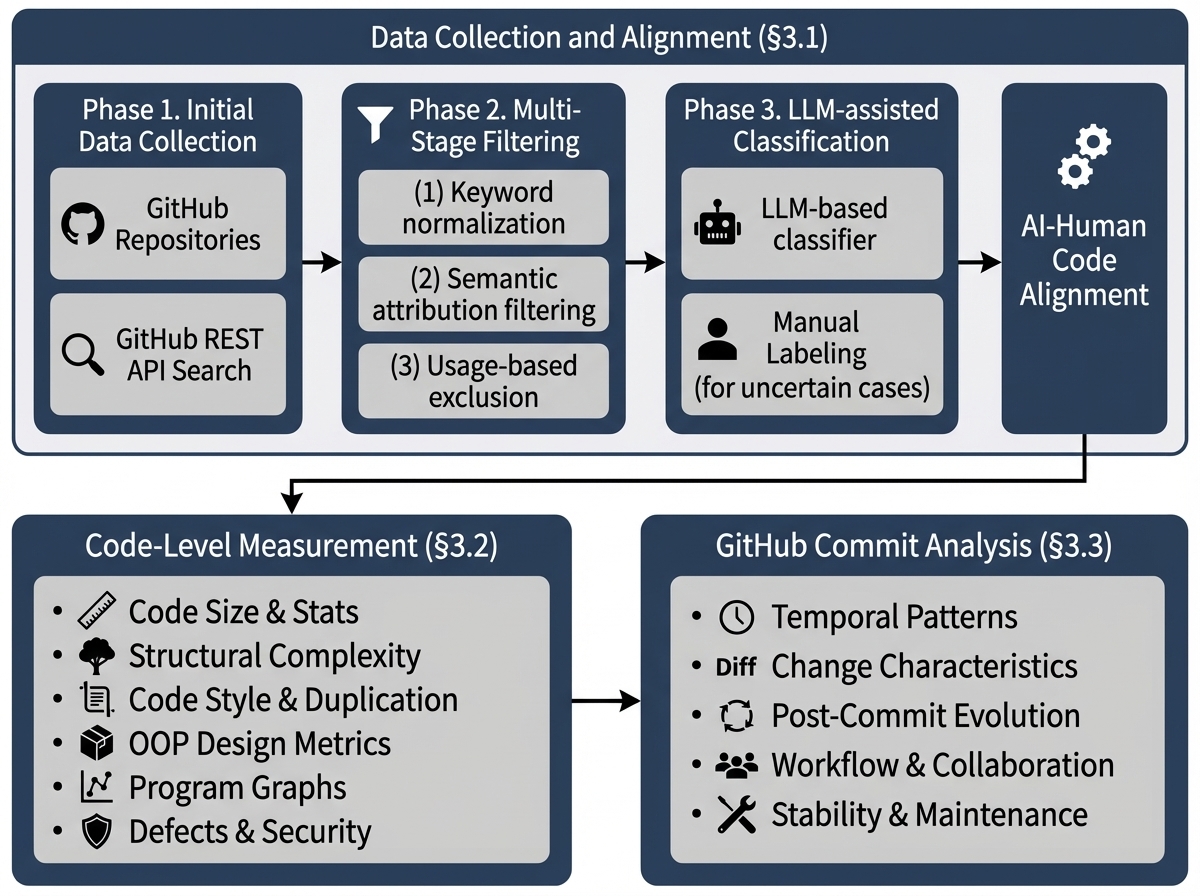}
    \caption{Measurement Pipeline.}
    \label{fig:pipeline}
\end{figure}

This section describes our measurement pipeline, from dataset collection to our considered metrics and measurement methodology. 
Figure~\ref{fig:pipeline} overviews our pipeline.

\subsection{Dataset Collection and Alignment}
\label{sec:dataset}

We first differentiate two concepts: \emph{AI-generated code} and \emph{AI-involved code}.
We say the code is AI-generated if everything is indeed produced by AI,
while AI-involved is weaker:
AI edits may be mixed with human coding (\eg some functions in a file are generated by AI and some written by human, that file thus becomes AI-involved).
We can similarly define \emph{human-written code} and \emph{human-involved code}.
As our goal is to understand the AI-human differences,
ideally,
we should only compare pure AI-generated code to human-written code.
However,
accurately setting apart AI-generated code within complex real-world codebases is difficult if not impossible,
given the frequently mixed developments and lack of clear indicators.
This also forms a core challenge for our study.
In this section,
we describe our procedure to obtain AI code that is \emph{closer} to AI-generated than AI-involved,
and human code that is \emph{closer} to human-written than human-involved.

\subsubsection{Collect AI-involved Code Files}
\label{sec:collect-ai-involved-file}

We start by collecting \emph{AI-involved} code files from real-world codebases.
Those are source files mixing AI and human coding.
To that end, we got inspiration from recent work~\cite{Schreiber_2025} that demonstrates the feasibility of identifying AI-involved code from GitHub by searching for explicit hints in code comments.
Such hints are combined using 6 prefix words (\ie by, with, use, used, using, from) and specific LLM tools (\eg ``by ChatGPT'').
However, this approach suffers from several limitations.
First, keyword search often introduces substantial noises due to substring matching (\eg matching ``use'' in unrelated words).
Second, attribution comments are frequently ambiguous, as mentions of LLMs may refer to non-coding tasks (\eg documentation or data generation).
Third, prior filtering strategies remain coarse-grained and cannot reliably distinguish true AI-involved code from incidental references.
To address these challenges, we design a refined data collection pipeline that improves both the precision and reliability:

\noindent\textbf{Phase 1. Initial Data Collection.}
We \emph{extend} the LLM tool list used in previous work~\cite{Schreiber_2025} \textit{(Chatgpt, Copliot, Tabnine, Codewhisperer)} to additionally include Llama, Claude, Gemini, DeepSeek, Grok, Qwen, Cursor AI, totally 11 LLM tools. This yields 66 search keywords after combining the 6 prefixes which we use to search with GitHub REST APIs. We also \emph{extend} the repository date range up to Dec 31, 2025. After applying \cite{Schreiber_2025}'s original filtering method, we collected 44,616 unique code files mentioned in 31,726 commits.

\noindent\textbf{Phase 2. Multi-Stage Filtering.}
Beyond the pipeline of \cite{Schreiber_2025},
we develop a multi-stage rule-based filtering procedure to further improve the data quality:

\noindent\textit{(1) Keyword normalization.}  
We eliminate false matches caused by substring-based search with whole-word matching for prefixes (\eg ``use'', ``by''), reducing noises introduced by the GitHub search mechanism.

\noindent\textit{(2) Semantic attribution filtering.}  
We identify explicit attribution patterns (\eg ``generated by'', ``written by'', ``enhanced by'') and analyze their surrounding context to ensure that the LLM is used for \emph{code generation}. Mentions referring to non-code artifacts (\eg text, media) are excluded.

\noindent\textit{(3) Usage-based filtering.}  
We remove cases where LLMs are mentioned only as tools for auxiliary tasks (\eg API calls, analysis, or communication) instead of code generation.

To implement (2) and (3),
we curated a list of keywords and phrases for the inclusion/exclusion patterns and encode them in a script.
The long list is omitted in the paper while the script will be released.
After this filtering, we obtain 8,268 AI-involved code files mentioned in 4,882 commits.
We then randomly sample 100 files from the 8,268 and the filtered-out ones, respectively, for a manual check regarding whether they are truly AI-involved or not.
The result shows a high accuracy (TP=88/100, TN=95/100), indicating the effectiveness and reliability of our rule-based filtering.

\noindent\textbf{Phase 3. LLM-assisted Classification.}
Not all code files can be classified by rules in Phase 2. For those cases, we employ an LLM-based classifier (ChatGPT 5-mini) to decide AI-involvement. This yields an additional 10,932 AI-involved files related to 7,877 commits. Remaining uncertain cases are manually labeled, resulting in 683 additional AI-involved files. A secondary manual sample check (similar to last phase) confirms the reliability of LLM-based decision (TP=85/100, TN=93/100).

We further deduplicate the collected files using repository, file path, and commit hash. Our final dataset contains \textbf{19,816} AI-involved code files related to \textbf{12,749} commits - the largest to our knowledge. Fig.~\ref{fig:tool_dist} and Fig.~\ref{fig:lang_dist} characterizes this dataset regarding its LLM and programming language distributions, respectively. Consistent with prior observations~\cite{Schreiber_2025}, general-purpose models (\eg ChatGPT) dominate real-world usage, while code spans a diverse set of languages with Python and JavaScript being the most prevalent.


\begin{figure}[!t]
\centering
\small

\begin{subfigure}{\columnwidth}
\centering
\begin{tikzpicture}
\begin{axis}[
    width=0.98\columnwidth,
    height=1.9cm,
    xbar stacked,
    xmin=0, xmax=100,
    axis x line*=bottom,
    axis y line*=none,
    ytick=\empty,
    xtick={0,20,40,60,80,100},
    xticklabels={0\%,20\%,40\%,60\%,80\%,100\%},
    tick style={draw=none},
    bar width=12pt,
    enlarge y limits=0.7,
    legend style={
        at={(0.5,1.45)},
        anchor=south,
        legend columns=3,
        draw=none,
        font=\small,
        /tikz/every even column/.append style={column sep=0.6cm}
    }
]

\addplot+[draw=black, fill=blue!60]   coordinates {(83.3,0)};
\addplot+[draw=black, fill=cyan!70]   coordinates {(10.8,0)};
\addplot+[draw=black, fill=yellow!60] coordinates {(2.2,0)};
\addplot+[draw=black, fill=orange!70] coordinates {(1.7,0)};
\addplot+[draw=black, fill=red!65]    coordinates {(1.3,0)};
\addplot+[draw=black, fill=blue!35]   coordinates {(0.7,0)};

\legend{
ChatGPT (83.3\%),
Copilot (10.8\%),
DeepSeek (2.2\%),
Gemini (1.7\%),
Cursor (1.3\%),
Others (0.7\%)
}
\end{axis}
\end{tikzpicture}
\caption{Distribution of AI-generated code by tools}
\label{fig:tool_dist}
\end{subfigure}

\vspace{0.8em}

\begin{subfigure}{\columnwidth}
\centering
\begin{tikzpicture}
\begin{axis}[
    width=0.98\columnwidth,
    height=1.9cm,
    xbar stacked,
    xmin=0, xmax=100,
    axis x line*=bottom,
    axis y line*=none,
    ytick=\empty,
    xtick={0,20,40,60,80,100},
    xticklabels={0\%,20\%,40\%,60\%,80\%,100\%},
    tick style={draw=none},
    bar width=12pt,
    enlarge y limits=0.7,
    legend style={
        at={(0.5,1.52)},
        anchor=south,
        legend columns=3,
        draw=none,
        font=\small,
        /tikz/every even column/.append style={column sep=0.5cm}
    }
]

\addplot+[draw=black, fill=blue!60]    coordinates {(41.7,0)};
\addplot+[draw=black, fill=cyan!70]    coordinates {(15.4,0)};
\addplot+[draw=black, fill=yellow!60]  coordinates {(8.6,0)};
\addplot+[draw=black, fill=orange!70]  coordinates {(7.8,0)};
\addplot+[draw=black, fill=red!65]     coordinates {(7.8,0)};
\addplot+[draw=black, fill=blue!35]    coordinates {(4.8,0)};
\addplot+[draw=black, fill=teal!35]    coordinates {(3.8,0)};
\addplot+[draw=black, fill=brown!20]   coordinates {(3.7,0)};
\addplot+[draw=black, fill=magenta!35] coordinates {(6.4,0)};

\legend{
Python (41.7\%),
JavaScript (15.4\%),
Java (8.6\%),
C++ (7.8\%),
TypeScript (7.8\%),
C\# (4.8\%),
Swift (3.8\%),
C (3.7\%),
Others (6.4\%)
}
\end{axis}
\end{tikzpicture}
\caption{Distribution of AI-generated code by languages}
\label{fig:lang_dist}
\end{subfigure}

\caption{Distribution of AI-generated code records by tools and programming languages.}
\label{fig:tool_lang_dist}
\end{figure}

\subsubsection{Extract Likely AI-generated Code Units}
\label{sec:collect-ai-gen-units}

AI-involved code files collected in \S\ref{sec:collect-ai-involved-file} may mix AI and human contributions. In this stage our goal is to extract code units that are \emph{likely} AI-generated. Our observation is that the AI-hinting comments used in \S\ref{sec:collect-ai-involved-file} are usually scoped to certain code units (\eg a function, a class). For example, a large file may contain an AI-hinting comment saying that ``the below function is generated by ChatGPT``. That comment is still insufficient to qualify the whole file as AI-generated, however the specific function at below is confidently AI-generated. It is also common that the AI-hinting comment is scoped to the whole file (\eg placed at the start of the file), indicating that the entire file is AI-generated. These observations lead to our high-level design: we will first identify the \emph{scope} of any single AI-hinting comment with a AI-involved file, then treat the code units within the scope as \emph{likely} AI-generated. Next, we provide details about our comment scope identification.

\noindent\textbf{Scope Definition.}
We consider common code scopes used in different programming languages.
This includes files,
functions (\eg methods, constructors, lambdas, \etc), 
and type declarations (\eg classes, interfaces, structs, namespaces, \etc).
For each supported scope, 
we extract the syntax rule to identify its boundary from the corresponding language specifications (\eg a \texttt{C} function starts with a prototype header and ends with the closing brace),
which supports our later scope identification.

\noindent\textbf{Comment-Scope Association.}
The goal of this step is to infer the finest code scope to which an AI-hinting comment applies. 
We begin by expanding the comment into a complete comment block, merging adjacent comment lines that are separated only by blank lines. 
We then associate the comment block with nearby syntactic scopes according to its location and directional cues.
Specifically,
if the comment block appears inside an existing function-like scope, we treat the smallest enclosing scope as the default target. 
However, when the comment is adjacent to a nested function-like or type-like declaration, we prefer the adjacent nested scope because such comments often describe the immediately following or prefixing code unit. 
By default, we select the following nested target (\eg a comment describing the overall semantic of a whole function usually appears just before the function);
we select the prefixing target only when the comment explicitly refers to preceding code using directional words such as \emph{above}, \emph{previous}, \emph{prior}, \emph{preceding}, \emph{earlier}, \emph{before}, or \emph{aforementioned}.

If the comment block is outside all function-like scopes, we determine whether it acts as a file-level header/footer or a local standalone comment. 
To identify header and footer regions, we ignore non-logical lines such as blank lines, comments, imports/includes, package or using declarations, decorators/annotations, and explicit constant declarations. 
A comment in the leading header region is interpreted as applying to the following file content: we extract all subsequent function-like targets, or the remaining executable top-level script if no function-like target exists. 
Similarly, a comment in the trailing footer region is interpreted as applying to the preceding file content: we extract all previous function-like targets, or the preceding top-level script when no function-like target exists. 
For standalone comments in the middle of a file, we select the next function-like target by default, unless the comment contains explicit backward-reference cues, in which case we select the previous target.

Finally, multiple AI-hinting comments in the same file may select overlapping scopes. We deduplicate overlapping targets by repository, file version, and source range. Following this procedure, we analyzed AI-involved code files from \S\ref{sec:collect-ai-involved-file}, extracted \textbf{42,792} likely AI-generated sub-file code units (\eg functions) and \textbf{9,843} likely AI-generated file snapshots.

\noindent\textbf{Validation.}
To assess whether comment-scope association indeed produces likely AI-generated code units, we manually inspected two random samples: 100 comment-associated functions and 100 comment-associated files. For each sample, we examined the AI-hinting comment, its surrounding context, the extracted code unit to determine if such scope truly described as AI-generated. The inspection shows 87/100 function-level units and 98/100 file-level units are true positives. This result supports that our scope-localization step substantially improves the precision of AI-generated-code identification.

\subsubsection{Human Code Alignment}
\label{sec:human-align}

After extracting likely AI-generated code units, we construct matched human controls for both function-level and file-level comparison.
This step serves two purposes: obtaining code that is closer to human-written code (instead of just human-involved), and aligning each human control with its AI counterpart to reduce confounding factors.
A direct comparison between all AI and all non-AI code may be biased by irrelevant factors like repository-specific coding styles, project maturity, functionality, code size, \etc
Therefore, we apply the following alignment criteria.

\noindent\textbf{Pre-AI and stable-period controls.}
To obtain controls that are more likely human-written code, we first select candidate controls from before the repository's earliest observed AI-involved commit, reducing the chance of unmarked AI assistance.
However, very old code may introduce temporal confounders, since early repository code can be immature or written under different conventions.
We therefore further restrict controls to a stable development period before the first observed AI-involved commit.
Intuitively, a stable period is a consecutive sequence of commits after the project has reached a mature size and is no longer rapidly expanding or shrinking.
Operationally, we use source-file count as the default stability metric and source lines of code as a fallback.
A stable period is a maximal consecutive interval with at least two commits, where every commit reaches at least 80\% of the final observed repository size and the metric variation within the interval is no more than 20\% of the final observed size.

\noindent\textbf{Repository, language, path, and size matching.}
For each likely AI-generated target, we search human controls from the same repository and same programming language, controlling for project-level practices and language-specific metric differences.
Within the repository, candidate controls are searched by path distance, starting from the same directory and gradually expanding to nearby directories and broader subtrees, so that controls are more likely to come from related project components.
We also apply size matching: a control must be within 20\% lines of code of the AI target; for targets with no more than five lines, we allow controls up to twice the size. If no sufficient controls are found under this strict constraint, we use a relaxed closest-size fallback.

\noindent\textbf{Final matched dataset.}
We attempt to select up to 2 human controls for each AI sample. If fewer than 2 controls are available but one valid control exists, we retain the AI sample with single matched control; samples without any valid control are excluded from matched analysis. After this preprocessing, \textbf{36,855} likely AI-generated sub-file units and \textbf{7,471} file snapshots are successfully aligned with stable pre-AI human controls. The final matched human-control dataset contains \textbf{65,391} likely human-written sub-file units and \textbf{15,424} files.

\subsection{Study Code-Level Characteristics}
\label{sec:code-method}

We aim at comprehensively understanding the differences between AI-generated and human-written code in real-world settings,
to that end, we measaure an extensive and diverse set of \emph{code-level characteristics},
surpassing prior works and providing a holistic comparison.

\noindent\textbf{Metric Taxonomy.}
Specifically, our scope includes the following categories of metrics:

\noindent\textit{(1) Code Size and Basic Statistics.}  
Basic properties like lines of code (LOC), number of files, token-level statistics, providing a baseline understanding of code scale and granularity.

\noindent\textit{(2) Structural Complexity.}  
We analyze fine-grained structural components extracted from abstract syntax trees (ASTs), including functions, parameters, classes, statements, conditionals, loops, and recursion. We further derive complexity-related metrics such as cyclomatic complexity, nesting depth, and decision counts.

\noindent\textit{(3) Code Style and Duplication.}  
Stylistic properties such as code duplication, reflecting maintainability and reuse patterns.

\noindent\textit{(4) Object-Oriented Design Metrics.}  
For object-oriented languages (\eg Java, C++, Python), we measure classical OOP metrics (\eg class-level complexity and coupling) to understand design quality.

\noindent\textit{(5) Program Graphs.}  
We analyze program characteristics from their control-flow, data-flow, and call graphs,
capturing higher-level properties regarding both syntax and semantic.

\noindent\textit{(6) Code Defects and Security.}  
We measure security-related properties such as vulnerability patterns and counts of potential defects,
gaining insights of comparative code quality.

\noindent\textbf{Metric Normalization.}
As the absolute volume of AI-generated and human-written code is different in our dataset (\S\ref{sec:dataset}), direct metric comparison by raw counts is misleading.
To address this, we normalize each metric using an appropriate \emph{measurement unit}.
For example, defect-related metrics are normalized by lines of code (\eg bugs per KLOC),
while structural metrics are measured at the function or file level (\eg average number of loops per function). 

\noindent\textbf{Statistical Significance Testing.}
To draw meaningful and reliable conclusions in code characteristic differences betwene AI and human,
we perform statistical hypothesis testing to determine whether observed AI/human differences are statistically significant.

\noindent\textbf{Implementation.}
We employ various tools to compute the aforementioned metrics (\eg \textit{cloc} for basic statistics, \textit{CK} and \textit{Understand} for OOP metrics, \textit{CodeQL} for security analysis, \etc).
More details can be found in Table~\ref{tab:code_metrics},
which summarizes all measured metrics, along with their analysis tools and normalization scopes.
\begin{table*}[t]
\centering
\small
\caption{Summary of code-level metrics, analysis tools, and measurement scopes.}
\label{tab:code_metrics}
\resizebox{0.97\textwidth}{!}{
\begin{tabular}{llll}
\toprule
\textbf{Category} & \textbf{Metrics} & \textbf{Tool} & \textbf{Scope} \\
\midrule

Code Size \& Statistics 
& LOC, token count, file size 
& cloc~\cite{adanial_cloc}, pygments~\cite{pygments} 
& file / function \\

Structural Complexity 
& \#functions, parameters, statements, conditions, loops, recursion 
& tree-sitter~\cite{treesitter} 
& per function / per file \\
& Cyclomatic complexity, nesting depth, decision count 
& AST-based analysis 
& function \\

Code Style 
& Code duplication ratio 
& jscpd~\cite{jscpd}
& file / project \\

OOP Design 
& Class/method metrics (e.g., complexity, coupling) 
& CK (Java)~\cite{cktool}, Understand (C++, Python)~\cite{understand} 
& class \\

Program Graphs 
& CFG/DFG properties, call graph size, dependencies 
& joern~\cite{joern} + networkx~\cite{networkx2008} 
& function / file \\

Security 
& Vulnerability count, CWE patterns 
& CodeQL~\cite{codeql} 
& per KLOC \\
& Hard-coded secrets (passwords, API keys) 
& String analysis 
& file \\

\bottomrule
\end{tabular}
}
\end{table*}
\subsection{Study Commit-Level Patterns}
\label{sec:commit-method}

Beyond code-level characteristics (\S\ref{sec:code-method}), it is equally important to understand how AI-generated code is \emph{produced, integrated, and evolved} within real-world development workflows.
Commit-level analysis provides a natural lens to study such dynamics, as commits capture both developer activity and project evolution.
By examining commit histories associated with AI-generated code, we aim to uncover how developers collaborate with AI tools in practice and how AI-assisted code influences development processes and outcomes.

\noindent\textbf{Research Questions.}
We specifically aim at answering the following research questions during our commit-level study,
which collectively characterize the \emph{real-world usage patterns and lifecycle dynamics} of AI-assisted code.

\begin{itemize}[leftmargin=*]
    \item \textit{RQ1: Temporal patterns.} Do AI-involved commits exhibit different temporal characteristics (e.g., working hours, release proximity) compared to human-written code?
    \item \textit{RQ2: Change characteristics.} Do AI-involved commits involve different magnitudes or types of changes (e.g., number of modified files, lines of code, churn)?
    \item \textit{RQ3: Post-commit evolution.} Are AI-involved commits more likely to be revised, reverted, modified after submission?
    \item \textit{RQ4: Collaboration and workflow.} How does AI-involved code affect developer collaboration patterns, such as handoffs and multi-author contributions? 
    \item \textit{RQ5: Stability and maintenance.} Does AI-generated code require more frequent fixes or exhibit different stabilization patterns compared to human-written code?
\end{itemize}

\noindent\textbf{Methodology.}
To answer these questions, we extract commit-level information from real-world repositories for both AI-associated and human-control commits. Specifically, we collect information from the following dimensions:

\begin{itemize}[leftmargin=*]
\item \textit{Temporal Characteristics (for RQ1).}  
Commit timestamps, including time of day, day of week, and distance to the nearest release.

\item \textit{Change Characteristics (for RQ2).}  
The scale and scope of code changes, including the number of modified files, lines, and overall churn.

\item \textit{Post-Commit Evolution (for RQ3, RQ4, RQ5).}  
The frequency of post-commit modification and revert within fixed time windows (7, 30, and 90 days), the number of subsequent edits, participating authors, time to first modification.

\item  \textit{Workflow and Maintenance Patterns (for RQ4, and RQ5).}  
These include \emph{fix likelihood}, determined by whether subsequent commits contain fix-related keywords (\eg ``fix'', ``revert''),  \emph{stabilization time}, measured as the time until the last modification of a file, and \emph{handoff latency}, defined as the time until another developer modifies the code. We consider related issues in the interval of 90 days after the commit.
\end{itemize}

\noindent\textit{Statistical Hypothesis Testing.}
Similar to \S\ref{sec:code-method}, we ensure that our comparative conclusions are statistically significant. In addition to pooled group-level tests, we conduct paired comparisons between each AI-associated commit and its matched human-control commit from the same repository. We use Chi-squared tests~\cite{Pearson01071900} for categorical variables and Mann-Whitney U tests~\cite{aoms/1177730491} for numerical variables in group-level analysis, and McNemar's tests~\cite{McNemar_1947} and Wilcoxon signed-rank tests~\cite{wilcoxon} for paired comparisons. For numerical paired comparisons, we report rank-biserial correlation $R_{rb}$\cite{Cureton_1956} as an effect size, where positive values indicate larger AI-associated values and negative values indicate larger human-control values.

\noindent\textbf{Implementation.}
To collect the aforementioned information and answer our research questions,
we mainly use the \texttt{git} toolset with various corresponding commands (\eg \texttt{git diff} for \textit{change characteristics} and \texttt{git log} for some temporal metadata, \etc).
\section{Code-Level Measurement Results}
\label{sec:code-measure}
\begin{figure*}[!t]
    \centering
    \includegraphics[width=\linewidth]{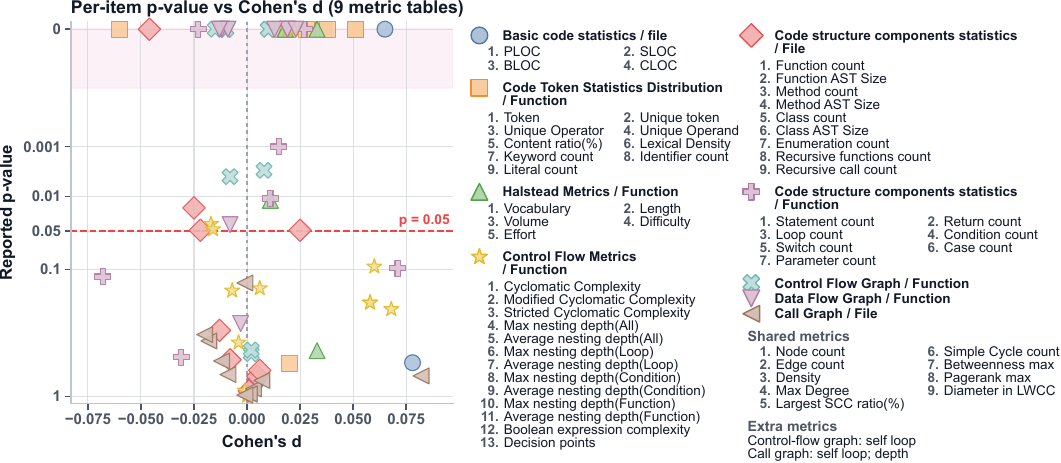}
    \caption{Statistic result of code structure metrics}
    \label{fig:code_strucure}
\end{figure*}

We present our measurement results and distilled findings regarding code-level metrics in this section.
The measurement is based on our matched dataset between likely AI-generated code and their human-written counterparts (\S\ref{sec:human-align}),
covering both sub-file code units and files.
As outlined in \S\ref{sec:code-method}, we aim to understand the AI-human differences on 77 code-level metrics across different categories.
We report the p-values and Cohen's d effect sizes for all code strcutural metrics in Figure~\ref{fig:code_strucure},
and OOP-related metrics in Figure~\ref{fig:code_oop}.

\subsection{Code Structure Statistics}
\paragraph{Code Length}
Comparing the lines of code (LOC) between AI-generated and human-written files,
we observed no statistically significant differences on PLOC, SLOC, and CLOC.
BLOC difference is statistically significant,
but the Cohen’s d value suggests a quite small effect size. 

\paragraph{Token Metrics}
We implemented token-based analysis at function level, including the amount of (unique) tokens, operators, operands, keywords, identifiers, content ratio and lexical density.
Based on these metrics, we also calculated the Halstead complexity of AI-generated functions and human-written ones. 
From Figure~\ref{fig:code_strucure}, AI-generated code is statistically different from human-written code (except for literal count and effort in Halstead metrics),
but their differences are again very small as suggested by the low Cohen’s d values.

\paragraph{Code Structure Metrics}
We next examine structural and control-flow complexity at both function and file levels, including file-level structures, function-level structures, control flow complexity metrics, graph structures of code mentioned in \S\ref{sec:code-method}. Results show that part of the metrics mentioned above are not statistically different between AI and human code, such as \texttt{case} amount, nesting depth of condition and functions, boolean expression and decision points, method, class, recursive call metrics, call graph metrics. Even for metrics with statistically significant differences, Figure~\ref{fig:code_strucure} still shows small Cohen’s d values for them, indicating tiny differences.

This result is surprising. 
Although we examine a substantially broader set of code-level metrics than prior work, we do not observe practically large differences between likely AI-generated and matched human-written code. 
As shown in Figure~\ref{fig:code_strucure}, many metrics are statistically significant, but their Cohen's \(d\) values remain small, indicating limited practical separation between the two groups.
This contrasts with prior studies that report more pronounced differences between AI-generated and human-written code in smaller-scale or more controlled settings~\cite{10.1109/ICSE55347.2025.00005,xu-etal-2026-code,rahman2025,ISSRE11229706}. 
One possible explanation is that real-world AI-assisted code is not raw model output: developers may review and iterate on AI suggestions (\eg via multi-round prompting and refinement) to fit the surrounding codebase, thereby reducing stylistic and structural gaps from existing human-written code.
\ibox{ins:code-structure}{AI-generated code is structurally similar to human code}{Although many structural metrics are statistically different, the corresponding effect sizes are small, suggesting that real-world AI-generated and human-written code are largely similar in structural complexity.}

\begin{figure}[!t]
    \centering
    \includegraphics[width=\linewidth]{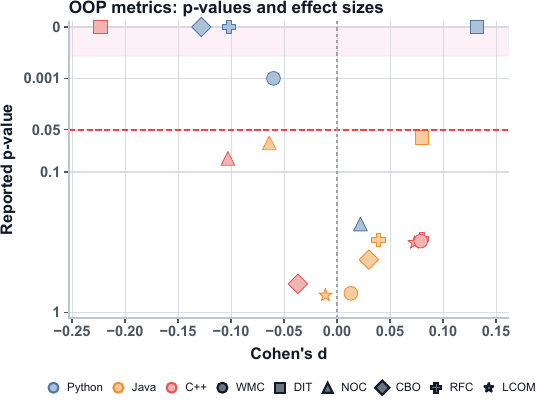}
    \caption{Statistic result of OOP metrics}
    \label{fig:code_oop}
\end{figure}

\paragraph{OOP Metrics}
We also analyze object-oriented design characteristics using standard CK-style metrics as listed in Figure~\ref{fig:code_oop}.
Results are reported per class across C++, Java, and Python.
A key observation is that the object-oriented design differences between AI and human code vary substantially across languages.
In C++, the notable difference is observed in DIT, where human-written code has deeper inheritance hierarchies than AI-generated code (d=-0.223). Nevertheless, this effect remains small, and the absolute DIT values are low for both groups. 
By contrast, none of JAVA OOP metrics shows a statistically significant difference, and all effect sizes are negligible.
Python falls between these two cases: human-written Python classes have slightly higher WMC, CBO, and RFC, suggesting marginally greater behavioral complexity and object coupling, while AI-generated Python classes exhibit a slightly higher DIT.
However, the corresponding Cohen’s d values range only from 0.060 to 0.132, indicating that these differences have limited practical significance.
Taken together, these results indicate that AI-generated and human-written code exhibit broadly similar object-oriented structures, with only minor language-specific deviations.
\ibox{ins:OOP}{OOP differences are language-specific but practically small}{AI-generated and human-written classes exhibit broadly similar object-oriented structures although some differences are statistically significant.}

\subsection{Code Style}
\label{sec:res-code-style}

Code style reflects how developers organize, format, and reuse code. We examine two complementary aspects: \emph{duplication patterns} and \emph{formatting structure}.

\begin{figure}[!t]
    \centering
    \includegraphics[width=\linewidth]{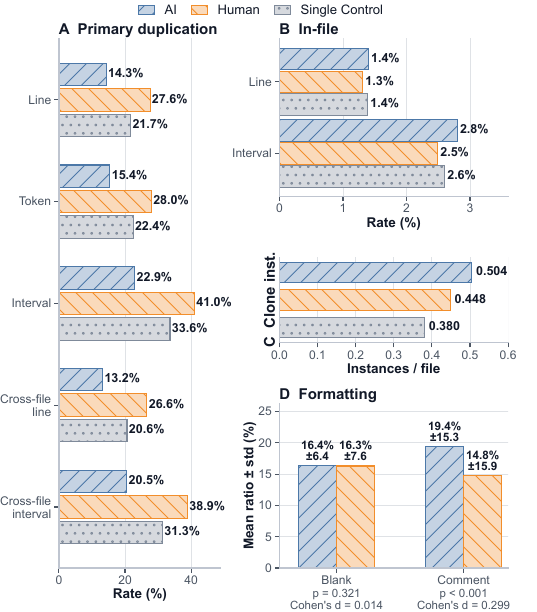}
    \caption{Code style and duplication metrics}
    \label{fig:code_duplicate}
\end{figure}

\paragraph{Ratio of Blank lines and comments}
The comment ratio metric in Figure~\ref{fig:code_duplicate} shows that AI tends to write more comments than human, supported by both p-value and Cohen’s d value.  

\ibox{ins:comment}{AI-generated code contains more comments}{AI-generated and human-written files show nearly identical blank-line ratios, however AI-generated files have a higher comment ratio, indicating that AI-generated code tends to include more explanatory text.}

\paragraph{Code Duplication}
Figure~\ref{fig:code_duplicate} summarizes duplication metrics. Overall, human-written code exhibits substantially higher duplication ratios than AI-generated code across all major measures (\eg duplicated lines: 27.57\% vs. 14.29\%, duplicated intervals: 41\% vs. 22.93\%). This difference is primarily driven by \emph{cross-file duplication}, where human-written code shows significantly higher reuse (26.62\% vs. 13.22\%). In contrast, within-file duplication remains low and similar across both groups. Interestingly, AI-generated code exhibits a slightly higher number of clone instances per file (0.504 vs. 0.448), suggesting a different duplication pattern.

Since we select up to 2 human-written controls for each AI sample, the human-control set may have a higher chance of cross-file duplication. We also repeated the analysis on a single-control subset that keeps only the first matched control for each AI sample. The result remains consistent that human-written code still shows a higher cross-file duplication rate than AI-generated code.

\ibox{ins:duplicate}{Human-written code exhibits more cross-file duplication}{Human-written code shows substantially higher duplication rates than AI-generated code, mainly due to cross-file duplication rather than within-file clones.}

\subsection{Code Security and Defects}
\label{sec:res-codeql}

We analyze security and defect characteristics using CodeQL and manual inspection of sensitive information leakage. Metrics are normalized per k-lines to ensure fair comparison.

\begin{figure}[!t]
    \centering
    \includegraphics[width=\linewidth]{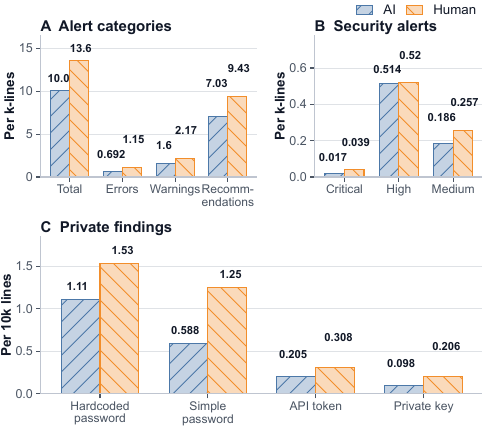}
    \caption{Overall CodeQL alerts and hardcoded secret findings}
    \label{fig:code_security_overall}
\end{figure}

\begin{figure}[!t]
    \centering
    \includegraphics[width=\linewidth]{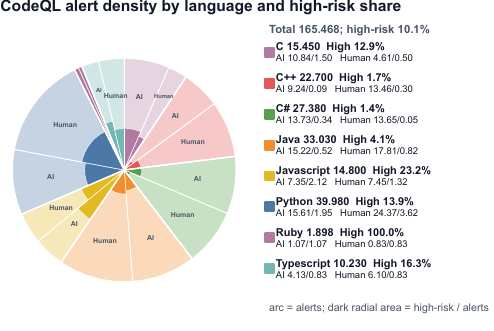}
    \caption{CodeQL alert density by language}
    \label{fig:code_security_language}
\end{figure}

\begin{figure}[!t]
    \centering
    \includegraphics[width=\linewidth]{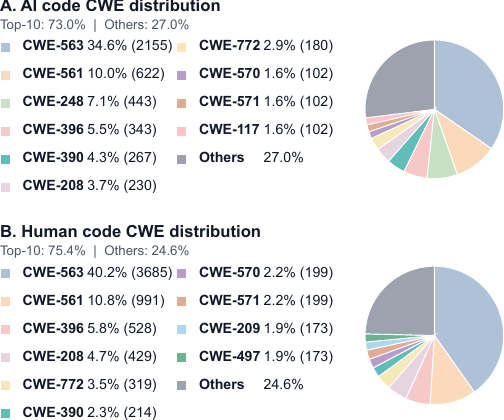}
    \caption{CWE composition of CodeQL alerts}
    \label{fig:code_security_cwe}
\end{figure}

\paragraph{Overall security and defect density}
As shown in Figure~\ref{fig:code_security_overall}, AI-generated code triggers less alerts than human-written code (10.04 vs. 13.56 per KLOC) in each type of alerts and each level of severity except high risk alerts that AI-generated code is similar with human-written code(0.514 vs. 0.52). 
This result contrasts with several prior studies \cite{ISSRE11229706,SMC10394237,mohsin2024,wang2024} suggesting that AI-involved code integrated into realistic development projects may exhibit different security patterns from raw model outputs.
\paragraph{Language-dependent security patterns}
Figure~\ref{fig:code_security_language} reveals strong language-specific effects. AI-generated code shows higher alert density in C and similar alert density with human-written code in C\# and JavaScript, while human-written code is higher in other languages. For high-risk alerts, C, C\#, JavaScript, and Ruby show higher densities in AI-generated code, whereas C++, Java, and Python show the opposite trend; TypeScript shows similar densities across the two groups.
\paragraph{Vulnerability composition}
Figure~\ref{fig:code_security_cwe} shows that both groups share the same dominant CWE categories (CWE-563, CWE-561), suggesting that unused or ineffective code constructs are the primary sources of CodeQL warnings in both groups. These two categories are more concentrated in human-written code than in AI-generated code (51.0\% vs. 44.6\%). Beyond these dominant categories, the distributions begin to diverge. AI-generated code shows relatively higher proportions of CWE-248 and CWE-390, indicating more exception-handling-related issues, and CWE-117 also appears among its top categories. In contrast, human-written code shows higher proportions of CWE-208 and CWE-772, and its top-10 list additionally includes CWE-209 and CWE-497. Overall, the results suggest that the two groups share similar leading vulnerability categories, but differ in their secondary issue composition: AI-generated code is relatively more associated with exception-handling and logging-related warnings, whereas human-written code contains relatively more concurrency/resource-lifecycle and information-exposure-related categories.
\paragraph{Sensitive information leakage}
Figure~\ref{fig:code_security_overall} shows that human-written code contains more hardcoded sensitive information overall, including passwords, API keys, tokens, and private keys.
\ibox{ins:security}{AI-generated code shows lower overall defect density, but security differences are language-dependent}{AI-generated code triggers fewer CodeQL alerts and contains fewer hardcoded sensitive values overall than matched human-written code. However, high-risk alerts vary substantially across languages, and both groups share the same dominant CWE categories.}

\section{Commit-Level Measurement Results}
\label{sec:commit-measure}
\begin{figure}[!t]
    \centering
    \includegraphics[width=\linewidth]{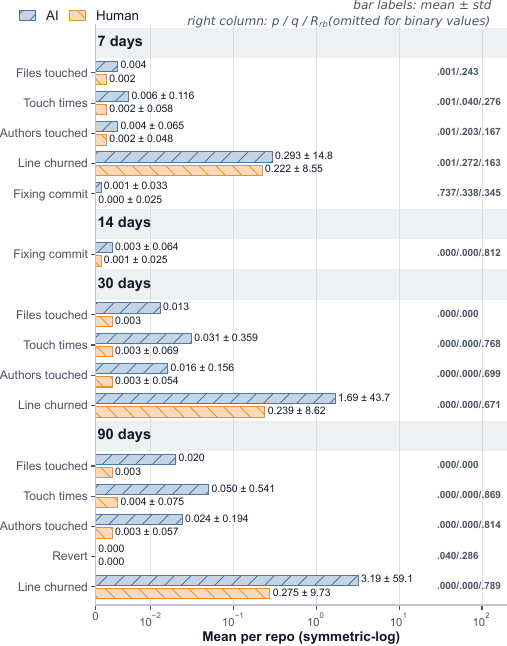}
    \caption{Post-commit evolution by timeline}
    \label{fig:commit1}
\end{figure}

\begin{figure}[!t]
    \centering
    \includegraphics[width=\linewidth]{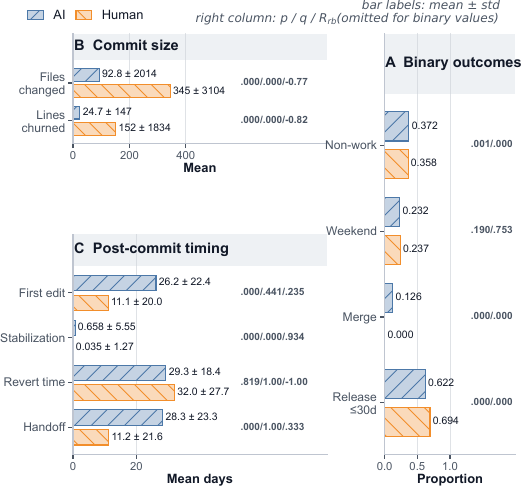}
    \caption{Commit size, timing, and binary commit outcomes}
    \label{fig:commit2}
\end{figure}

Figure~\ref{fig:commit1} and Figure~\ref{fig:commit2} summarize commit-level comparisons between AI-involved and human-written code, and we report both pooled group-level and matched pair-wise results. The \emph{$p$-value} compares all AI-involved samples with all human-control samples as two pooled groups. The \emph{$q$-value} is the FDR-adjusted $p$-value from matched pair-wise comparisons between each AI-involved sample and its matched human-control commit. When two controls are available, we use the first matched control for commit-level analysis. Throughout this paper, values below \emph{0.05} are considered statistically significant. Therefore, significance in both \emph{$p$} and \emph{$q$} indicates a robust difference; significance only in \emph{$p$} indicates a group-level difference that weakens after matching; significance only in \emph{$q$} indicates a difference that is clearer after matching; and significance in neither indicates no reliable difference. $R_{rb}$ is the rank-biserial effect size and sign indicates direction for non-binary values. Results are organized by the research questions introduced in \S\ref{sec:commit-method}.
\paragraph{RQ1: Temporal patterns}
Commits containing AI-involved code are slightly more likely to occur during non-working hours (0.372 vs. 0.358), while weekend activity shows no consistent difference under paired comparisons. In addition, commits with AI-involved codes are less likely to occur within 30 days before a release (0.622 vs. 0.694).
\ibox{ins:offhour-support}{Flexible but Risk-Aware Usage}{
Commits with AI-involved codes are more likely to occur outside regular working hours, but are less frequent near release deadlines, suggesting that developers use AI flexibly while avoiding it in stability-critical stages.
}
\paragraph{RQ2: Change characteristics}
AI-involved commits are substantially smaller in scope. They involve fewer files and significantly lower line churn per commit compared to human-written code.
\ibox{ins:localized-commit}{Localized and Fine-Grained Edits}{
AI-assisted coding is primarily used for small, localized modifications rather than large-scale, multi-file changes, indicating a role as a fine-grained development aid.
}
\paragraph{RQ3 \& RQ5: Post-commit evolution and maintenance}
Differences are limited in the short term (7 days), but become pronounced over longer horizons. AI-involved code exhibits more follow-up activity within 30 and 90 days, with higher touch frequency, more participating authors, and greater accumulated churn. It also shows slightly higher likelihood of follow-up fixing commits and significantly longer stabilization time. However, the low mean values suggest that these effects are concentrated in a subset of commits with repeated follow-up activity. This finding differs from \cite{xiao2026}, which measures post-commit activity within a 14-day window, whereas our analysis extends the horizon to 90 days and reveals longer-term follow-up patterns.
\ibox{ins:delayed-maintenance}{Delayed but Concentrated Maintenance}{
AI-involved code does not show substantially higher immediate instability, but incurs more follow-up modifications and longer stabilization over time, suggesting increased downstream adjustment costs concentrated in a subset of cases.
}
\paragraph{RQ4: Collaboration and workflow}
AI-involved commits involve longer handoff latency (28.28 vs. 11.23 days) and more follow-up contributors over time, indicating a different collaboration pattern. At the same time, merge commits appear more frequently in AI-involved changes.
\ibox{ins:workflow-shift}{Slower Handoffs and Distributed Follow-Up}{
AI-involved code tends to remain longer with the original author before being handed off, but eventually involves more distributed follow-up activity, suggesting a shift toward delayed but broader collaboration.
}
\section{Discussion}
\label{sec:discussion}

\noindent\textbf{Scope and Interpretation.}
Unlike prior work that evaluates LLM-generated code under controlled prompts, we study the \emph{observable outcome of real-world AI-assisted programming}: code that has been selected, edited, and integrated into real repositories. Our goal is not to characterize all possible model outputs, but to understand the properties of \emph{AI-involved code in practice}. Since AI involvement is rarely explicitly labeled in the wild, we rely on the multi-stage filtering and validation pipeline described in \S\ref{sec:dataset}. Thus, our results should be interpreted as large-scale trends over high-fidelity samples rather than claims about every individual code fragment.

Under this scope, our findings suggest that AI-involved code is not radically different from matched human-written code in structural metrics; many observed differences are statistically significant but practically small. The more consistent differences appear in style, reuse, security patterns, and development process: AI-involved code is more comment-heavy, less cross-file reused, exhibits language-dependent security differences, and is introduced through smaller, more localized commits that require more follow-up modification over longer horizons. These results reflect the combined effect of model behavior and human integration, rather than raw model outputs alone.

\vspace{3pt}
\noindent\textbf{Implications.}
For developers, our results suggest that AI assistance reshapes development effort rather than simply reducing it. AI-involved code appears mainly in localized edits, but may shift part of the effort from initial coding to downstream validation and maintenance. Developers should therefore allocate review and testing effort beyond the immediate post-commit window, especially for code that continues to receive follow-up changes.

For AI tool vendors, our findings suggest that generation-time assistance should be complemented with stronger post-generation support. Future tools could better help developers maintain cross-file consistency, identify code that may require later stabilization, and provide language-aware recommendations, given the substantial language-dependent differences observed in security and defect patterns.

\vspace{3pt}
\noindent\textbf{Methodological Considerations.}
Several factors should be considered when interpreting our findings. First, although we match samples within the same repository, language, and similar code size, residual differences in development context, contributor behavior, and project maturity may remain. Second, our measurements depend on static analysis, code metric, and clone detection tools, whose coverage and precision vary across languages. We therefore emphasize recurring cross-metric trends rather than isolated metric-level differences. Third, because our dataset is large, small effects can become statistically significant. We therefore interpret results using both statistical significance and effect size, and use paired analyses where applicable.
\section{Conclusion}
This paper presents a large-scale empirical study of likely AI-generated code in real-world repositories, comparing code selected and integrated into real projects with matched human-written code. We find that such code is similar to human-written code in structural metrics, but more comment-heavy, less cross-file reused, exhibits language-dependent security differences. At the process level, AI-involved commits are smaller and more localized, yet require more follow-up modification over time. These results suggest that AI assistance serves as a local development aid while shifting effort toward downstream validation and maintenance. Our released pipeline enables future studies to track this evolving practice.

\Urlmuskip=0mu plus 1mu
\bibliographystyle{IEEEtran}
\bibliography{Reference}
\end{document}